\documentclass[aps,twocolumn,english,superscriptaddress,citeautoscript,preprintnumbers,amsmath,amssymb,floatfix,footinbib]{revtex4-1}
\usepackage{graphicx}
\usepackage{graphics}
\usepackage{bm}
\usepackage{amssymb }
\usepackage{array}
\usepackage[breaklinks=true,colorlinks,citecolor=blue,linkcolor=blue,urlcolor=blue]{hyperref}
\usepackage{float}
\usepackage[table]{xcolor}
\usepackage{cancel}

\definecolor{Cyan}{rgb}{0.4,1,0.7}
\definecolor{LightCyan}{rgb}{0.8,1,0.85}

\begin{document}

\title{Topological Hall effect in the antiferromagnetic Dirac semimetal EuAgAs}

\author{Antu Laha}
\email{antuiitk2012@gmail.com}
\affiliation{Department of Physics, Indian Institute of Technology, Kanpur 208016, India}

\author{Ratnadwip Singha}
\thanks{Present address: Department of Chemistry, Princeton University, Princeton, New Jersey 08544, USA}
\affiliation{Saha Institute of Nuclear Physics, HBNI, 1/AF Bidhannagar, Calcutta 700 064, India}

\author{Sougata Mardanya}
\affiliation{Department of Physics, Indian Institute of Technology, Kanpur 208016, India}

\author{Bahadur Singh}
\email {bahadur.singh@tifr.res.in}
\affiliation{Department of Condensed Matter Physics and Materials Science, Tata Institute of Fundamental Research, Mumbai 400005, India}

\author{Amit Agarwal}
\affiliation{Department of Physics, Indian Institute of Technology, Kanpur 208016, India}

\author{Prabhat Mandal}
\affiliation{Saha Institute of Nuclear Physics, HBNI, 1/AF Bidhannagar, Calcutta 700 064, India}

\author{Z. Hossain}
\email{zakir@iitk.ac.in}
\affiliation{Department of Physics, Indian Institute of Technology, Kanpur 208016, India}

\begin{abstract}
The non-trivial magnetic texture in real space gives rise to the intriguing phenomenon of topological Hall effect (THE), which is relatively less explored in topological semimetals. Here, we report large THE in the antiferromagnetic (AFM) state in single crystals of EuAgAs, an AFM Dirac semimetal. EuAgAs hosts AFM ground state below $T_N$ = 12 K with a weak ferromagnetic component. The in-plane isothermal magnetization below $T_N$ exhibits a weak metamagnetic transition. We also observe chiral anomaly induced positive longitudinal magnetoconductivity which indicates a Weyl fermion state under applied magnetic field. The first-principles calculations reveal that EuAgAs is an AFM Dirac semimetal with a pair of Dirac cones, and therefore, a Weyl semimetailic state can be realized under time-reversal symmetry breaking  via an applied magnetic field. Our study establishes that EuAgAs is a novel system for exploiting the interplay of band topology and the topology of the magnetic texture.
\end{abstract}

\maketitle

The rich interplay between the electronic structure and topology in materials ascertain a variety of topological states. These include topological states as diverse as  Dirac or Weyl semimetals (DSMs or WSMs), topological insulators, topological magnets, and topological superconductors \cite{RevModPhys_WSMs,Review_WSMs,RevModPhys_TIs,RevModPhys_TIs_SCs,SmB6_NatureCom_2013,YbCdGe_PRB_2019,CaCdSn_PRB_2020,YbCdSn_PRB_2020,Co3Sn2S2_Scince_2016,Co2MnGa_science_2019,Co3Sn2S2_Nature_phys_2018,Co3Sn2S2_Nature_comn_2018,TbPtBi_PRB_2020,Weyl_magnon_PRL_2016, Weyl_magnon_PRL_2017, Weyl_magnon_natcomn_2016, Weyl_magnon_natphy_2018}. The topological states of matter are robust and cannot be destroyed by local disorder and perturbations. Among these, the topological magnets are particularly appealing since they support tunable phases and electromagnetic responses which are exciting for fundamental science as well as for next-generation technological applications. Notable examples are magnetic WSMs where magnetization can enhance the net Berry flux through Weyl nodes tuning and as a consequence,  the anomalous Hall effect (AHE) is enhanced \cite{Co3Sn2S2_Nature_phys_2018, Co3Sn2S2_Nature_comn_2018,ZrTe5_Nature_phy_2018,GdPtBi_natphy_2016}. Despite their recent experimental realization, the current choice of the available materials is quite limited. Therefore, search for new topological magnets with novel electromagnetic responses is of immense importance.

Furthermore, topologically non-trivial spin textures in magnetic systems give rise to another type of Hall effect, termed as topological Hall effect (THE). Such nontrivial spin textures in real space are generated either by geometrical frustration or antisymmetric Dzylashinsky-Moriya (DM) interaction \cite{THE_PRL_2004}. When an electron moves through these spin textures, it acquires a real space Berry phase, which acts as an effective magnetic field.  This effective magnetic field is proportional to the scalar spin chirality of the magnetic structure and leads to THE \cite{Wang2019}. The nontrivial spin textures can be described by a finite topological number $Q=(1/4\pi) \int \hat{n} \cdot(\partial \hat{n}/\partial x \times \partial \hat{n}/\partial y)~dx dy$, where $\hat{n}$ is the unit vector in the direction of local magnetization. An integer value of $Q$, with $|Q|\geq 1$, results in THE \cite{MnSi_PRL_2009,MnGe_PRL_2011,Schulz2012,THE_biskyrmion,MnPdGa_APL_2019}. Recent studies have shown that achiral polar crystals with $C_{nv}$ symmetry allow DM interaction which can stabilize the Neel-type skyrmion phase with large THE \cite{CeAlGe_PRL_2020,VOSe2O5_PRL_2017,GaV4S8_nmat_2015}. While the THE has been observed in various families of materials as diverse as B20 compounds, antiferromagnets, pyrochlore lattices, correlated oxide thin films \cite{Mn5Si3_Nature_comn_2014,UCu5_Nature_comn_2012,Spin_Chirality_Science_2001,Pr2Ir2O7_PRL_2007,EuO_PRB_2015,CaCeMnO3_Nature_phys_2019,MnSi_PRL_2009,MnGe_PRL_2011,FeGe_PRL_2012,AuFe_PRL_2004,AuFe_PRB_2006}, it remains largely unexplored in topological WSMs with a few exceptions such as CeAlGe \cite{CeAlGe_PRL_2020}.

In this Letter, we report large THE in single crystalline EuAgAs, a new antiferromagnetic (AFM) DSM. EuAgAs crystallizes in a hexagonal symmetry with $P6_3/mmc$ ($D_{6h}$) space group. This symmetry group can generate finite scalar magnetization-chirality as well as skyrmion or bi-skyrmion phases in magnetic materials, suitable for realizing the THE \cite{THE_biskyrmion, MnPdGa_APL_2019}. We explore the interplay between the topology and magnetism in EuAgAs below $T_N=$12 K. We also observe chiral anomaly induced positive longitudinal magnetoconductivity associated with the Weyl fermion state under applied magnetic field. From the first-principles calculations, we conclude that EuAgAs realizes an AFM DSM state with a pair of Dirac cones lying on the $C_{3z}$ axis. Thus, our study establishes EuAgAs as a new AFM DSM with anomalous transport properties.

\begin{figure}
	\centering
	\includegraphics[width=0.99\linewidth]{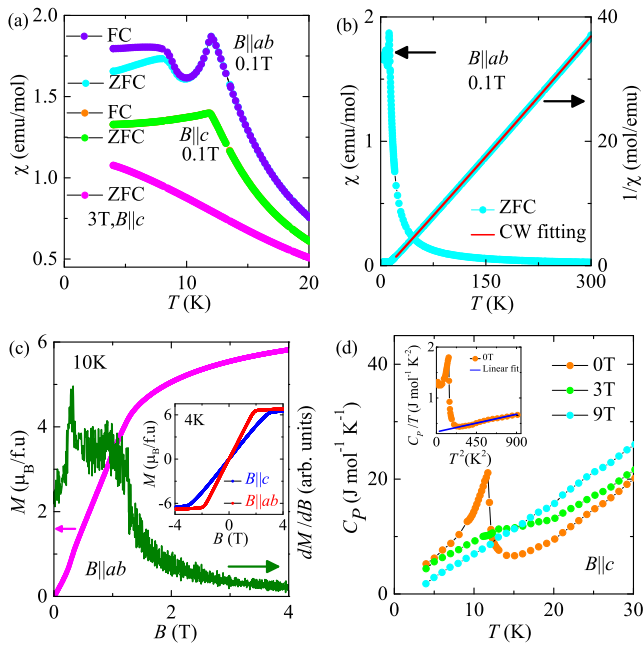}
	\caption{(a) Temperature-dependent magnetic susceptibility in zero-field-cooled (ZFC) and field-cooled (FC) conditions for $B$$\parallel$$c$ and $B$$\parallel$$ab$. (b) The ZFC magnetic susceptibility (cyan markers) and the Curie-Weiss fitting (solid red line) to the $1/\chi$ curve in the temperature range $20-300$ K for $B$$\parallel$$ab$. (c) Isothermal magnetization ($M$) and the first derivative of $M$ ($dM/dB$) at 10 K for $B$$\parallel$$ab$. The inset shows $M$ vs. $B$ at 4 K for $B$$\parallel$$c$ and $B$$\parallel$$ab$. (d) Temperature dependence of the specific heat at various magnetic field values up to 9T. The solid blue line in the inset fits the formula $C_P/T=\gamma+\beta T^2$.}
	\label{Fig1}
\end{figure}

EuAgAs single crystals were grown by flux method with bismuth as an external flux \cite{CaAgAs_TIs_IOP}. Magnetotransport, magnetization, and specific heat measurements were carried out in  physical property measurement system and SQUID-VSM (Quantum Design). The experimental and computational details are given in supplementary material (SM) \cite{SM,Fullprof_1993,hohenberg1964,kohn1965,blochl1994_PAW,kresse1996,kresse1999,perdew1996_gga,anisimov1997first, anisimov1997second}. To resolve the magnetic ordering in EuAgAs, we have measured the temperature-dependent zero-field-cooled (ZFC) and field-cooled (FC) susceptibility ($\chi$) along the crystallographic $c$-axis ($B$$\parallel$$c$) and $ab$-plane ($B$$\parallel$$ab$) as shown in Fig.~\ref{Fig1}(a). The $\chi$ exhibits a peak at 12 K due to AFM ordering of the Eu$^{2+}$ moments which is suppressed by the application of a magnetic field of 3 T. It is evident from Fig.~\ref{Fig1} that the nature of $\chi$($T$) curve for $B$$\parallel$$ab$ is quite different from typical AFM system \cite{Ce2Ni3Ge5_JMMM_2018}. The $\chi$($T$) shows nonmonotonic $T$ dependence below $T_N$. After a sharp drop below $T_N$, $\chi$($T$) starts to increase and saturates below 8 K. A small bifurcation between ZFC and FC curves is also observed below 8 K. This behavior of $\chi$ suggests the presence of a weak FM component  and competition between AFM and ferromagnetic (FM) interactions \cite{TbNi2B2C_PRB_1996}. Above 20 K, we have fitted the 1/$\chi$ for $B$$\parallel$$ab$ with the modified Curie-Weiss law, $\chi(T)=\chi_0+C/(T-\theta_p)$ [Fig.~\ref{Fig1}(b)]. Here, $\chi_0$, $C$, and $\theta_P$ are the temperature-independent susceptibility, Curie constant, and Curie-Weiss temperature respectively. The estimated effective magnetic moment of Eu$^{2+}$ is 8.03 $\mu_B$ for $B$$\parallel$$ab$ and 7.68 $\mu_B$ for $B$$\parallel$$c$, which are close to the theoretical value of $g\sqrt{S(S+1)}\mu_B=7.94 \mu_B$ for $S=7/2$. The fit yields  $\theta_P \approx$ 10.4 K and 8.7 K for $B$$\parallel$$ab$ and $B$$\parallel$$c$, respectively [see Supplementary Materials (SM)] \cite{SM}.

To obtain further insight into the magnetic ground state, we have measured isothermal magnetization $M(B)$ for both orientations $B$$\parallel$$ab$ and $B$$\parallel$$c$. For $T$$<$$T_N$, the $M$ increases faster with increase in $B$ and saturation-like behaviour appears at lower field for $B$$\parallel$$ab$ ($\sim$ 2 T) as compared to $B\parallel c$ ($\sim$ 3 T) [see the inset of Fig.\ref{Fig1}(c)]. The $M(B)$ curve at 10 K shows a weak anomaly below 1 T for $B$$\parallel$$ab$ as shown in Fig.\ref{Fig1}(c). To analyze this behavior, we plot $dM/dB$ as a function of $B$, in which a peak is observed around $\sim$ 0.3 T. At low temperature, the $M(B)$ curve for $B$$\parallel$$ab$ exhibits an upward curvature and a weak hysteresis at low field (see SM for details) \cite{SM}. These behaviors indicate a weak metamagnetic transition for $B$$\parallel$$ab$. However, no anomaly is observed for $B\parallel c$ at low field. It is also clear from Figs.~\ref{Fig1}(a) and (c) that the easy axis of magnetization is on the $ab$-plane. The observed behavior is qualitatively similar to that reported for isostructural EuCuAs compound \cite{EuCuAs_JAC_2014}. Both the value and $T$ dependence of $\chi$ below $T_N$ in EuCuAs are also sensitive to the direction of the applied magnetic field and $\theta_P$ is positive. To reconcile the observed behavior of magnetic properties, A-type AFM ordering was proposed for EuCuAs where Eu$^{2+}$ spins align ferromagnetically within $ab$ plane and antiferromagnetically between the planes. Our theoretical results discussed below also support such a spin state for EuAgAs. We have also measured field dependence of magnetization at a high temperature above $T_N$ as shown in SM \cite{SM}. Even well above $T_N$, the $M$ is large and shows nonlinear B dependence (e.g. at 20 K).  

The temperature dependence of specific heat ($C_P$) at various fields up to 9 T is shown in Fig.~\ref{Fig1}(d).  $C_P$($T$) shows a sharp $\lambda$-like peak at $T_N$=12 K which suppresses with increase in magnetic field. The fit to zero-field $C_P$($T$) curve in the range 16-30 K with the expression $C_P=\gamma T+\beta T^3$  yields $\gamma\sim 334$ mJ mol$^{-1}$ K$^{-2}$ [inset of Fig.~\ref{Fig1}(d)]. Large value of $\gamma$ has also been reported for EuFe$_2$P$_2$ and EuCr$_2$As$_2$ and was attributed to magnon contributions from the ordering of Eu moments\cite{EuFe2P2_PRB_2010, EuCr2As2_PRB_2014}.

\begin{figure*}
\centering
\includegraphics[width=0.99\linewidth]{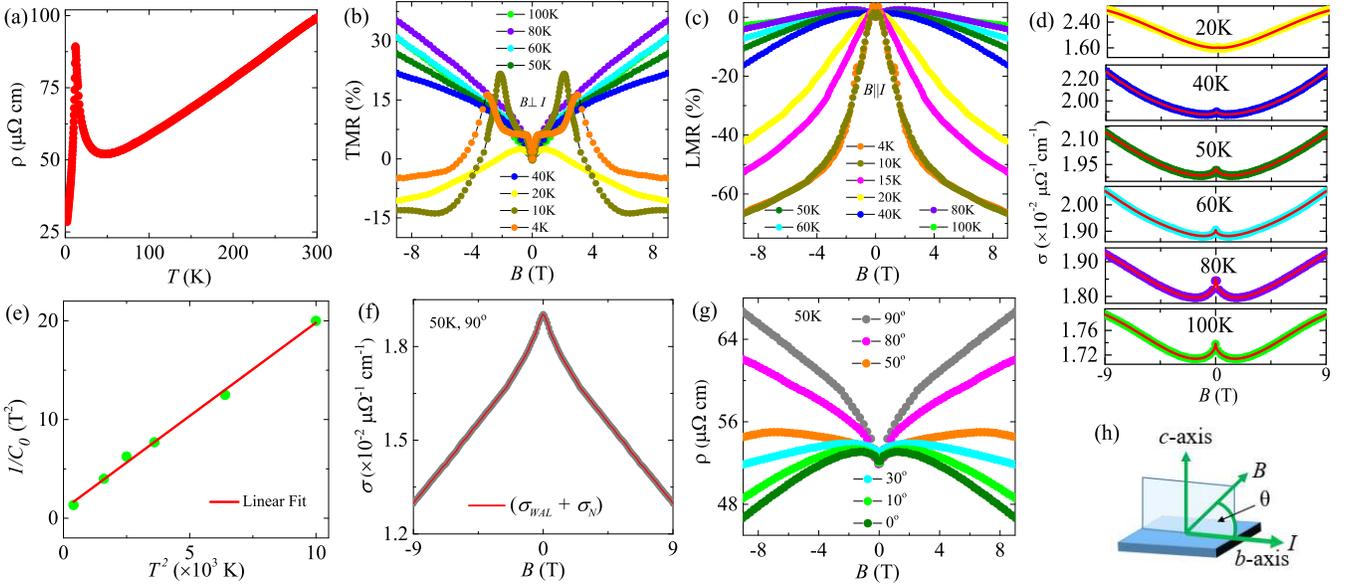}
\caption{(a) Temperature-dependent resistivity at zero magnetic field. (b) Transverse magnetoresistance ($B \perp I$) as a function of $B$ at different temperatures up to 100K. (c) Longitudinal magnetoresistance ($B \parallel I$) as a function of $B$ at various temperatures. (d) The magnetic field dependence of longitudinal conductivity ($\sigma=1/\rho$) at various $T$ ($\theta=0^\circ$). The solid red lines show the fit to Eq.~\eqref{Equ1}, which includes antilocalization and chiral anomaly induced corrections to conductivity. (e) $1/C_0$ vs. $T^2$, the red solid line indicates a linear fitting. (f) Transverse conductivity ($\sigma=\rho/(\rho^2+\rho_{yx}^2)$) for 50 K ($\theta=90^\circ$) fits to the expression $ \sigma(B)=\sigma_{WAL}+\sigma_N=(\sigma_0+a\sqrt{|B|})+(\rho_0+A B^2)^{-1}$. (g) Resistivity as a function of $B$ at various angle $\theta$ at 50 K. (h) Schematic diagram of the experimental set up.}
\label{Fig2}
\end{figure*}

We now discuss the electrical resistivity ($\rho$) and magnetoresistance (MR) of EuAgAs. Figure.~\ref{Fig2}(a) shows that $\rho$ decreases almost linearly with decreasing $T$ down to 40 K. With a further decrease in $T$ below 40 K, $\rho$ starts increasing due to the influence of magnetic ordering and reaches a maximum at $T_N$=12 K. The transverse magnetoresistance (TMR) shows a cusp-like feature at very low magnetic field which is attributed to the weak antilocalization (Fig.\ref{Fig2}(b)). The transition from the AFM state to a spin-polarized state is marked by a sharp maximum around 3 T. In the spin-polarized state, the TMR becomes negative and saturates at high fields. The TMR is negative up to 20 K. However, TMR becomes positive well above $T_N$ ($T>$ 20 K). In contrast, the longitudinal magnetoresistance (LMR) is negative both below and above $T_N$ and increases monotonically with the field, as shown in Fig.~\ref{Fig2}(c). The negative LMR increases continuously with a decrease in $T$ from 2.7\% at 100 K to 66\% at 4 K. The negative MR at low-temperature may partly arise due to the suppression of spin-disorder scattering as in the case of several Eu-based compounds where MR can be scaled with magnetization as discussed in SM~\cite{SM,Eu14MnBi11, Eu2CuSi3_PRB_1999}. However, the negative LMR in the paramagnetic region well above $T_N$ is surprising. Also, the significantly larger value of LMR and the nature of its field dependence suggest that LMR at low temperature is dominated by a nonmagnetic component. Regarding the physical mechanism behind the observation of the negative LMR, the possibility of the current jetting effect can be ruled out as discussed in the SM \cite{SM,Ag2Se_PRL_2005}. We attribute the negative LMR to the chiral anomaly associated with the presence of opposite chirality Weyl Fermions. This is supported by the successful fitting of the magnetoconductivity (MC) data to the semiclassical formula [Fig.~\ref{Fig2}(d)]~\cite{ABJ_anomaly_PRL_2013},
\begin{equation}
\sigma(B) = \sigma_{\rm WAL}~(1+C_0 B^2) + \sigma_N~.
\label{Equ1}
\end{equation}
Here, $\sigma_{\rm WAL}=(\sigma_0+a\sqrt{|B|})$ is the conductivity due to weak anti-localization effect, and $\sigma_N=(\rho_0+A B^2)^{-1}$ represents the normal conductivity originating from the conventional Fermi surface (other than Weyl points). $\rho_0$ denotes the zero-field resistivity, and $a$ and $A$ are two constants. The positive MC (or negative MR) depends on $C_0 B^2$, where the positive value of $C_0$ originates from the finite ${\bf E}\cdot{\bf B}$ term which produces chiral charge current in WSM~\cite{PhysRevResearch.2.013088,Nishihaya_PRB_2018, Ishizuka_PRB_2019}. Remarkably, we observe chiral anomaly and weak antilocalization up to relatively high temperature [Fig~\ref{Fig2}(d)] as in the case of YbPtBi \cite{YbPtBi_NatureCom_2018}. Notably, we do not find any $B-$linear resistivity terms in our fitting which indicates there are no electron pockets in the quantum limit \cite{Abrikosov_2000}. The temperature dependence of $C_0$ can be described by the relation $C_0\propto v_F^3 \tau_v/(T^2+\mu^2/\pi^2)$, where $v_F$, $\tau_v$ and $\mu$ are the Fermi velocity, chirality-changing scattering time and chemical potential respectively \cite{ZrTe5_Nat_Phys_2016}. The linear fitting to the curve $1/C_0$ vs. $T^2$ as shown in Fig.~\ref{Fig2}(e) further supports the presence of chiral anomaly in the temperature range 20-100 K. The estimated value of $\tau_v$ is $2.8 \times 10^{-12}$ s. Similar range of scattering timescales have also been seen in other Weyl semimetals \cite{TaAs_PRB_2020,YbPtBi_NatureCom_2018,TaAs_Nature_comn_2016}. The chiral anomaly induces positive MC only for the longitudinal case when $I \parallel B$, and this is not present in the transverse case (${\bf E}\cdot {\bf B}=0$). Notably, the transverse MC can be expressed as $ \sigma^T(B)=(\sigma_0+a\sqrt{|B|})+(\rho_0+A B^2)^{-1}$, without chiral anomaly. The successful fitting of MC data at 50 K to this expression further confirms that the origin of LMR and TMR are quite different and the cusp-like feature is due to the weak antilocalization [Fig.~\ref{Fig2}(f)]. We have further fitted the MC data with the Hikami-Larkin-Nagaoka formula to establish the weak antilocalization effect up to 100 K and shown the variation of phase coherence length with temperature in SM \cite{SM,Bi2Te3_PRB_2017,LuPdBi_SR_2014,LuPtSb_AIP_2015}.

Figure \ref{Fig2}(g) shows magnetic field dependent resistivity at various angle $\theta$ between $B$ and $I$ for 50 K (The direction of $B$ and $I$ are schematically shown in Fig.~\ref{Fig2}(h)). The MR shows a maximum negative value at $\theta=0^\circ$ (=-10$\%$) and decreases with increasing $\theta$. MR becomes positive for $\theta >30^\circ$ and reaches maximum at $\theta=90^\circ$ (=29$\%$). Similar angle-dependent behavior of the MR has also been reported in other WSMs \cite{Cd3As2_natcomn_2016, YbPtBi_NatureCom_2018}. We have also measured the directional dependence of MR by fixing the angle between $B$ and $I$ and observed a twofold symmetric anisotropy pattern at 4 K (details in SM) \cite{SM}.

\begin{figure}
	\centering
	\includegraphics[width=0.99\linewidth]{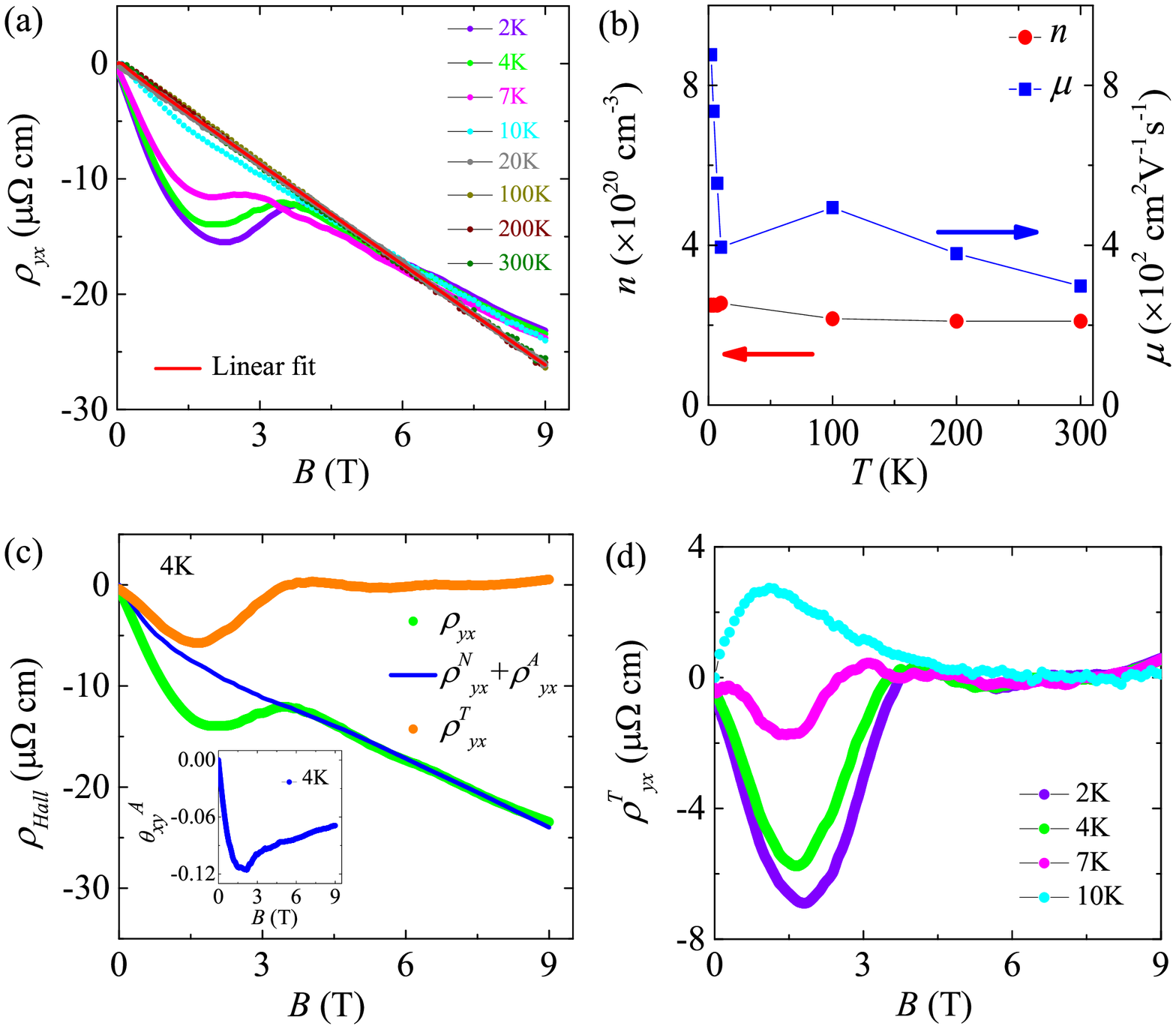}
	\caption{(a) The Hall resistivity $\rho_{yx}$ as a function of $B$ ($B \parallel c$) at various temperatures. (b) Charge carrier density ($n$) and mobility ($\mu$) vs.  temperature. (c) Hall resistivity (green markers) fitted to the conventional Hall contribution $\rho_{yx}$=$\rho^N_{yx} + \rho_{yx}^A$ at 4 K (blue solid lines). The clear deviation from the conventional Hall resistivity indicates the presence of an additional term - the topological Hall resistivity (orange markers). The inset shows Anomalous Hall angle as a function of $B$ at 4 K. (d) Topological Hall resistivity at various temperatures.}
	\label{Fig3}
\end{figure}

The experimental Hall resistivity after removing the MR contribution using the expression $\rho_{yx}=[\rho_{yx}(B)-\rho_{yx}(-B)]$ is shown in Fig.~\ref{Fig3}(a). The negative and linear $\rho_{yx}$ for 20 K $\leq T \leq$ 300 K indicates that electrons are the majority carriers. We estimate the carrier density ($n$) and Hall mobility ($\mu$) using the relations $n$=1/($eR_\textrm{0}$) and $\mu$=$R_\textrm{0}$/$\rho_{(B=0)}$, where $R_\textrm{0}$ is the slope of $\rho_{yx}(B)$ curve. The temperature dependence of $n$ and $\mu$ are shown in Fig.~\ref{Fig3}(b). We find that the carrier density is of the order of $10^{20}$cm$^{-3}$ which remains almost temperature independent.

The Hall resistivities for $T\leq$ 10 K show an anomaly below 3 T, which is quite prominent at low temperature. This behavior may be the reminiscent of the anomalous Hall effect (AHE) observed in AFM Weyl semimetals.  AHE is empirically modelled as $\rho_{yx}^A = S_0 \rho^2 M$, $S_0$ being a field independent parameter. However, we find that below 3 T, the observed $\rho_{yx}$  deviates from the fitting of the empirical Hall resistivity $\rho_{yx}=\rho_{yx}^N + \rho_{yx}^A$, as shown in Fig.~\ref{Fig3}(c). We also tried fitting the $\rho_{yx}$ with the conventional two-carriers model but it fails to reproduce the anomaly (details in SM) \cite{SM}. This clearly suggests the presence of an extra contribution in the Hall resistivity, which we attribute to the topological Hall effect \cite{MnGe_PRL_2011,CeAlGe_PRL_2020}.

By combining both AHE and THE in the AFM state, the total Hall resistivity can be expressed as \cite{MnGe_PRL_2011}

\begin{equation} \label{Eq2}
\rho_{yx}=\rho_{yx}^N + \rho_{yx}^A + \rho_{yx}^T = R_0B + S_0\rho^2 M + \rho_{yx}^T~
\end{equation}

Here, $\rho_{yx}^T$ denotes the novel topological Hall resistivity arising from the chiral magnetization-texture. Since $\rho_{yx}^T$ vanishes in the fully spin polarized state due to the absence of spin-chirality, we execute a linear fit to the curve $\rho_{yx}/B$ vs $\rho^2 M /B$ in the field range 4 T $\leq B \leq$ 9 T. From the fitting, we estimate, $R_0=-2.51~ \mu\Omega$ cm T$^{-1}$ and $S_0=-1.72 \times 10^{-3}$ ($\mu\Omega$ cm)$^{-1}$ $\mu_B^{-1}$. Using these parameters, $\rho_{yx}^T$ can be obtained by subtracting $\rho_{yx}^N + \rho_{yx}^A$ from $\rho_{yx}$, which is shown in Fig.\ref{Fig3}(c) \cite{CeAlGe_PRL_2020,MnGe_PRL_2011,MnSi_PRL_2013}. We estimate the maximum amplitude of the $\rho_{yx}^T$ at 2 K to be $\sim -7 ~\mu\Omega$ cm [Fig.\ref{Fig3}(d)]. This value of $\rho_{yx}^T$ is quite large. In order to compare the THE, we estimate the topological Hall angle, $\Theta_{xy}^T = \sigma_{xy}^T/\sigma$, where $\sigma_{xy}^T$ and $\sigma$ are the topological Hall conductivity and magnetoconductivity respectively. The maximum amplitude of $\Theta_{xy}^T$ ($\sim 0.17$) is comparable with that obtained in the magnetic Weyl semimetals GdPtBi ($\sim 0.17$) and YbPtBi ($\sim 0.18$) \cite{GdPtBi_natphy_2016,YbPtBi_NatureCom_2018}. We have also estimated the anomalous Hall angle ($\theta_{xy}^A=\sigma_{xy}^A/\sigma$) (inset of Fig.\ref{Fig3}(c)). The maximum amplitude of $\theta_{xy}^A$ is 0.11, which is compared with some other known AFM Weyl semimetals in Table-I of SM \cite{SM,Mn3Sn_Nature_2015,Mn3Ge_Science_adv_2016,GdPtBi_natphy_2016,TbPtBi_PRB_2019,TbPtBi_PRB_2020}.

\begin{figure}
\includegraphics[width=0.99\linewidth]{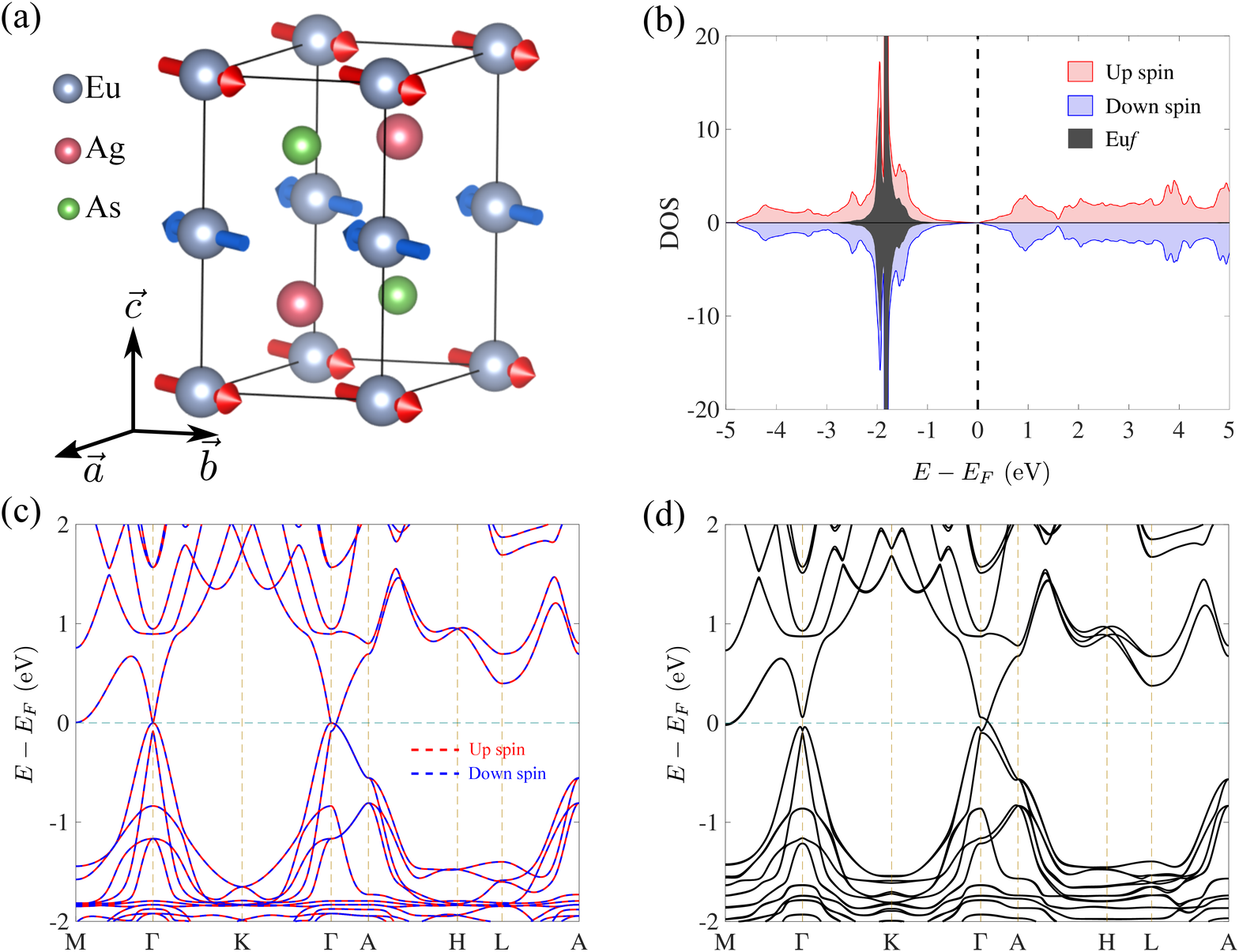}
\caption{ (a) Orientation of Eu spins in the antiferromagnetic EuAgAs. (b) The density of states (DOS). The partial DOS of spin-up and spin-down states are shown in red and blue, respectively. The DOS associated with the localized Eu $f$ states is shown in gray. (c) Band structure along the high-symmetry directions without spin-orbit coupling (SOC). The red and blue identify spin-up and spin-down states. (d) Band structure with SOC with spins along the $ab$-plane. The Dirac band crossings are seen along the $\Gamma$-$A$ direction.}
\label{DFT}
\end{figure}

We delineate the AFM DSM state of EuAgAs in Fig.~\ref{DFT}. The unit cell consists of six atoms, in which Eu atoms occupy the Wyckoff position 2a and form hexagonal stacking along the $z-$direction. The Ag and As atoms are placed at Wyckoff positions 2d and 2c respectively to form hexagonal layers which sandwiched between the Eu-layers, satisfying the overall inversion symmetry. To consider the possible magnetic configuration of EuAgAs, we investigated three AFM spin configurations (see SM for details) \cite{SM}. The lowest energy is obtained for the in-plane AFM [110] configuration as shown in Fig.~\ref{DFT}(a). This is consistent with our measured magnetic properties. The AFM ordering of the Eu $4f$ spins is further demonstrated in Fig.~\ref{DFT}(b) where the density of states (DOS) in the valence band region is dominated by the Eu $f$ states which is equal but opposite for spin-up and spin-down states. Moreover, EuAgAs has a vanishingly small DOS at the Fermi level which confirms its semi-metallic electronic state. This is further showcased in the band structure without spin-orbit coupling (SOC) in Fig.~\ref{DFT}(c). The valence and conduction bands derived for both the spin states are seen to cross along the $\Gamma-A$ direction with an inverted band ordering at the $\Gamma$-point. The band structure with SOC is shown in Fig.~\ref{DFT}(d). We find that the EuAgAs crystals have an inversion symmetry $I$, along with an  effective time-reversal symmetry, $S=\Theta \tau_{1/2}$, where $\Theta$ is the time-reversal operator and $\tau_{1/2}$ is translation vector connecting spin-up and spin-down atoms. Owing to these two symmetries, each of the band in AFM EuAgAs remains two-fold spin degenerate. We find that while the nodal band crossings without SOC are lifted, there are two symmetry-related four-fold degenerate band crossing points along the $\Gamma-A$ directions as shown in Fig.~\ref{DFT}(d). EuAgAs thus realizes a robust AFM DSM ground state with a pair of Dirac cones in vicinity of the Fermi energy, on the $C_{3z}$ rotation axis.

We next turn to discuss the possible reason behind the AHE and THE observed in EuAgAs. To confirm if the Dirac nodes in EuAgAs can split into a pair of Weyl nodes on applying a magnetic field, we calculated the electronic structure of the fully polarized state of EuAgAs (see SM) \cite{SM}. We find that each Dirac node indeed splits into a pair of Weyl nodes of opposite chiral charge, realizing a ferromagnetic Weyl semimetal. The distance between two pair of Weyl nodes is $k_{w1}$ = 0.1096 \AA$^{-1}$ and $k_{w2}$ = 0.227 \AA$^{-1}$. The calculated anomalous Hall conductivity using the expression $|\sigma^A_{yx}| = [(k_{w1}+k_{w2})/2\pi] \times (e^2/h)$ \cite{AHC_PRL_2011,Cd3As2_nano_PRB_2019,Smejkal2018,GdPtBi_Nature_mat_2016} is found to be 208 $\Omega^{-1}$cm$^{-1}$, which is lower than our experimentally observed maximum amplitude of $\sim 2700$ $\Omega^{-1}$cm$^{-1}$ at 2 K. The fact that EuAgAs belongs to a symmetry group which supports real space Berry curvature associated with nontrivial spin texture similar to MnNiGa and MnPdGa \cite{THE_biskyrmion, MnPdGa_APL_2019} makes THE a natural choice for the additional contribution.

In summary, using first-principle calculations we show that EuAgAs hosts an antiferromagnetic DSM ground state with a pair of Dirac cones. The observed magnetic properties confirm that EuAgAs shows an AFM transition at $T_N=$12 K along with a weak metamagnetic transition below $T_N$. Chiral anomaly induced positive longitudinal MC confirms the presence of a Weyl fermion state in EuAgAs under an applied magnetic field. It further exhibits anomalous and topological Hall effects. We provide compelling evidence of topological Hall effect in a spin only divalent europium compound that is likely to inspire further studies on topological Hall effect in DSMs.

{\it Acknowledgements.} Research support from IIT Kanpur is gratefully acknowledged. Z. H. also acknowledges support from SERB India (Grant No. CRG/2018/000220).  A. A. acknowledges Science and Engineering Research Board (SERB) and Department of Science and Technology (DST) of the government of India for financial support. The work at TIFR Mumbai is supported by the Department of Atomic Energy of the Government of India under project number 12-R\&D-TFR-5.10-0100. We thank Arup Ghosh for Hall resistivity measurements.

\bibliography{ReferenceAll}

\end{document}